\documentclass[showpacs,preprintnumbers,amsmath,amssymb]{revtex4}

\usepackage{graphicx}
\usepackage{dcolumn}
\usepackage{bm}
\begin{document}
\title{Quasielastic and inelastic neutrino reactions in $^{12}C$ at K2K energies}
\author{M. Sajjad Athar, S. Chauhan and S. K. Singh}
\affiliation{Department of Physics, Aligarh Muslim University, Aligarh-202 002, India}
\date{\today}
\begin{abstract}
In this paper, we present the results of a study made for the effect of nuclear medium in the charged current induced quasielastic lepton production(CCQE) and the incoherent and coherent one pion production (CC1$\pi^+$)processes from $^{12}C$ in the $\nu_\mu$ energy region of 0.4-3GeV. The theoretical results are compared with the recent experimental results for the ratio of charged current $\nu_\mu$ induced one pion production cross section to the quasielastic lepton production cross section reported by K2K collaboration. We also present the results for the angular and momentum distributions of leptons and pions produced in these processes.
\end{abstract}
\maketitle
\section{INTRODUCTION}
Recently K2K collaboration has reported experimental results on the inclusive single pion production induced by charged current by neutrinos in the energy region of 0.4-3 GeV\cite{Rodriguez}. Similar results were earlier reported by MiniBooNE collaboration in the energy region of 1GeV \cite{Miniboone}. The study of the energy dependence of the pion production cross sections in neutrino reactions for nuclei in this energy is important in modeling the neutrino nucleus cross sections for various Monte Carlo neutrino event generators used in analyzing the present neutrino oscillation experiment at MiniBooNE, K2K and future experiments to be done by T2K\cite{T2K} and NO$\nu$A \cite{Nova} collaborations.

The present experimental result by K2K collaboration reports the energy dependence of the ratio of inclusive cross sections of the charge current induced pion production and quasielastic reaction in neutrino reaction on polystyrene($C_8H_8$) target. Both of these reactions are influenced by the nuclear medium effects in the energy region of this experiment. The Monte Carlo simulation of single pion production in this reaction uses the Rein and Sehgal model\cite{Rein} which does not include any nuclear medium effect in the production process. The NEUT generator used in the analysis of K2K experiment\cite{Rodriguez}, however, includes the nuclear effects arising due to final state interactions of pions with the nucleus like pion absorption and pion scattering \cite{Neut}. The quasielastic reaction, on the other hand, uses Smith-Moniz model \cite{Moniz} which does not include the effect of nuclear medium arising due to nucleon-nucleon correlations but includes only the effect of Pauli principle and Fermi motion in a Fermi gas model.
There are many calculations for the nuclear medium effects in the quasielastic reaction~\cite{llewelyn}-\cite{Leitner}, and quite a few calculations in the case of single pion production which take into account explicitly the nuclear medium and final state interaction effects in various energy region~\cite{a2}-\cite{RusoCoh} . In this paper we present a study of nuclear medium effects in these reactions in the energy region of 0.4-3 GeV averaged over the neutrino flux of the K2K experiment.
The single pion production in this region is dominated by the resonance production in which a $\Delta$ resonance is excited and decays subsequently to a pion and a nucleon. When this process takes place inside the nucleus then there are two possibilities, the target nucleus remains in the ground state leading to coherent production of pions or is excited and/or broken up leading to incoherent production of pions. In the present paper, we have considered both the production processes in the $\Delta$ resonance model in the local density approximation to calculate single pion production from nuclei. The effect of nuclear medium on the production of $\Delta$ is treated by including the modification of $\Delta$ properties in the medium. Once pions are produced, they undergo final state interactions with the final nuclei. For the incoherent pion production final state interaction has been treated using a Monte Carlo code for pion nucleus interaction~\cite{Vicente}, while for the coherent case, the distortion of the pion plane wave is calculated by using the Eikonal approximation~\cite{Carrasco}. In the case of quasielastic reaction, the effects of Pauli principle and Fermi motion are included through the Lindhard function calculated in local density approximation~\cite{Singh1}. The details of these calculations are given in earlier publications~\cite{Athar2},\cite{a2},\cite{Athar4}.

 In this paper we report the results for the nuclear medium effects on total cross section $\sigma(E)$, $Q^2$ distribution $\frac{d\sigma}{dQ^2}$ and momentum distribution $\frac{d\sigma}{dp_l}$  for the muon, in the case of quasielastic and single pion production induced by charged currents. We compare our results of the ratio $R(E)=\frac{\sigma^{CC1\pi^{+}}(E)}{\sigma^{CCQE}(E)}$, with the recent observations of K2K collaboration. The results for the pion momentum and angular distributions have also been shown for the incoherent pion production process. In Secs. II \& III, we briefly describe the various formula used in the present calculations for quasielastic and inelastic reactions and present the results and discussion in Sec.IV.
\section{QUASIELASTIC REACTION}
The basic reaction for the quasielastic process is a neutrino interacting with a neutron inside the nucleus i.e.
\begin{equation}\label{quasi_reaction}
\nu_{\mu}(k) + n(p) \rightarrow \mu^{-}(k^{\prime}) + p(p^{\prime})
\end{equation}
The cross section for quasi-elastic charged lepton production is calculated in the local density approximation by taking into account the Fermi motion and the Pauli blocking effects through the imaginary part of the Lindhard function
for the particle hole excitations in the nuclear medium. The
renormalization of the weak transition strengths are calculated in the Random
Phase Approximation(RPA) through the interaction of the p-h excitations as
they propagate in the nuclear medium using a nucleon-nucleon potential
described by pion and rho exchanges~\cite{Oset1}. The effect of the Coulomb distortion
of muon in the field of final nucleus is also taken into account
using a local version of the modified effective momentum
approximation(MEMA)~\cite{Athar2}. 
\begin{figure}
\includegraphics{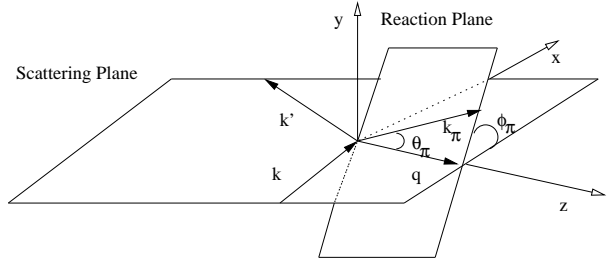}
\caption{Kinematical variables for $\nu_{\mu}(k) + N(p) \rightarrow \mu^{-}(k^{\prime}) + N^{\prime}(p^{\prime}) + \pi^{+}(k_{\pi})$ process.}
\end{figure}
The total cross section $\sigma(E_\nu)$ for the charged current neutrino induced reaction on a nucleon inside the nucleus in a local Fermi gas model is written as~\cite{Athar2}:
\begin{eqnarray}\label{sigma_quasi}
\sigma(E_\nu)&=&-\frac{2{G_F}^2\cos^2{\theta_c}}{\pi}\int_{r_{min}}^{r_{max}} r^2 dr \int^{{k^\prime}_{max}}_{{k^\prime}_{min}} {k^\prime}^2dk^{\prime} \int_{-1}^1dcos\theta \frac{1}{E_{\nu_\mu} E_\mu} L_{\mu\nu} J^{\mu\nu}_{RPA} Im{U_N(q_0, {\bf q})},
\end{eqnarray}
where $L_{\mu\nu}=\sum L_\mu {L_\nu}^\dagger$ and ${J^{\mu\nu}_{RPA}}={\bar\sum}\sum J^\mu {J^\nu}^\dagger$, calculated with RPA correlations in nuclei. 

The leptonic current $L_{\mu}$ and the hadronic current $J^\mu$ are given by
\begin{equation}\label{lep_curr}
L_{\mu}=\bar{u}(k^\prime)\gamma_\mu(1-\gamma_5)u(k)
\end{equation}
\begin{eqnarray}\label{had_curr}
J^\mu&=&\bar{u}(p^\prime)[F_{1}(q^2)\gamma^\mu + F_{2}(q^2)i{\sigma^{\mu\nu}}{\frac{q_\nu}{2M}} + F_{A}(q^2)\gamma^\mu\gamma_5 + F_{P}(q^2)q^\mu\gamma_5]u(p).
\end{eqnarray}
\begin{figure}
\includegraphics{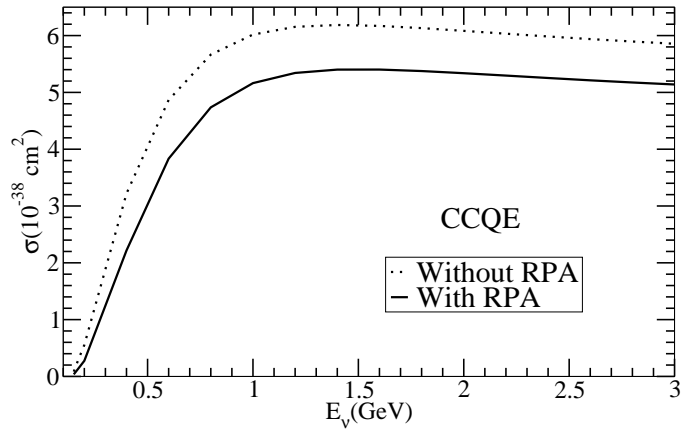}
\caption{Total scattering cross section for $\nu_{\mu}$ induced charged current quasielastic process in $^{12}C$.}
\end{figure}

\begin{figure}
\includegraphics{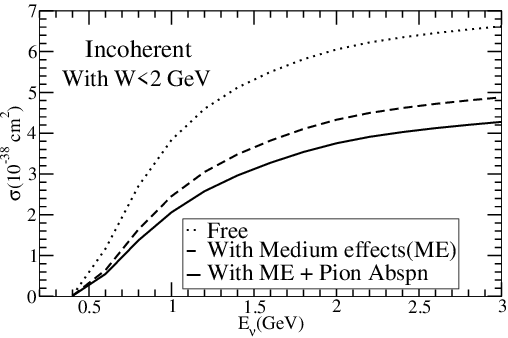}
\caption{Total scattering cross section for the $\nu_{\mu}$ induced incoherent charged current one pion production process in $^{12}C$.}
\end{figure}

\begin{figure}
\includegraphics{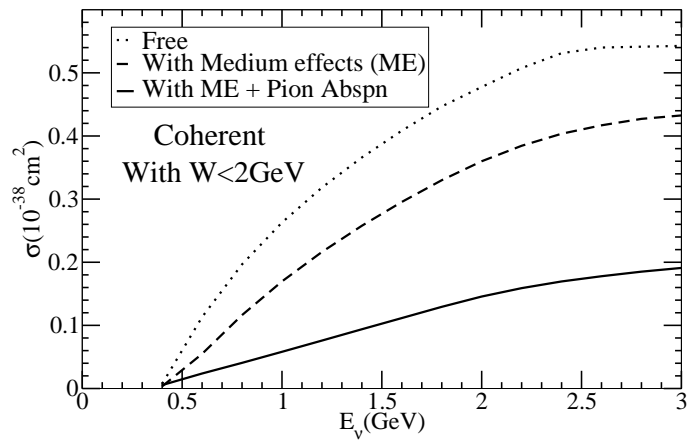}
\caption{Total scattering cross section for $\nu_{\mu}$ induced coherent charged current one pion production process in $^{12}C$.}
\end{figure}
\begin{figure}
\includegraphics{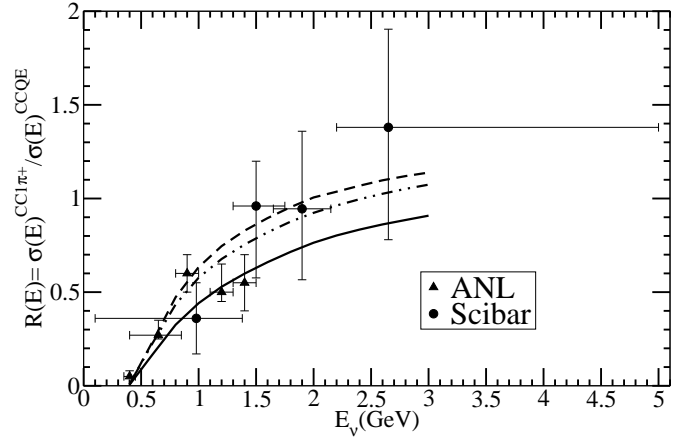}
\caption{Ratio of the cross sections for $\nu_{\mu}$ induced charged current one pion production process and charged current quasielastic process in polystyrene. Experimental points have been taken from Ref.\cite{Rodriguez} for K2K results in Scibar and from Refs.\cite{Radecky}, \cite{Barish} for ANL results.}
\end{figure}

\begin{figure}
\includegraphics{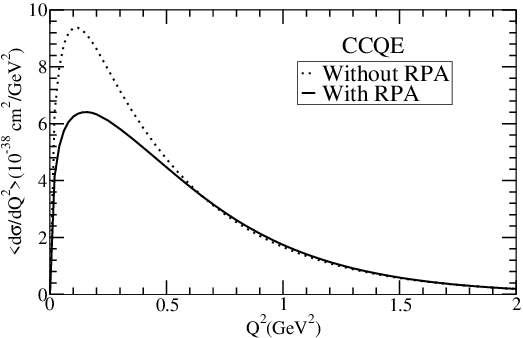}
\caption{$Q^2$ distribution for $\nu_{\mu}$ induced charged current quasielastic process in $^{12}C$ averaged over the K2K spectrum.}
\end{figure}

where $q(=k-k^\prime)$ is the four momentum transfer, M is the mass of the nucleon, $G_{F}(=1.16637\times 10^{-5} GeV^{-2})$ is the Fermi coupling constant and $\theta$ is the lepton angle. $U_N$ is the Lindhard function for the particle hole excitation~\cite{Singh1}. The form factors $F_1(Q^2)$, $F_2(Q^2)$, $F_A(Q^2)$ and $F_P(Q^2)$ are isovector electroweak form factors. For our numerical calculations we have used the parameterization of Bradford et al.~\cite{bradford} for the vector form factors $F_{1}(Q^2)$ and $F_{2}(Q^2)$ with dipole mass ${M}_{V}$=0.84GeV and a dipole form for $F_{A}(Q^2)$ with dipole mass $M_A$=1.1GeV.

Inside the nucleus, the Q-value of the reaction and Coulomb distortion of outgoing lepton are taken into account by modifying the Lindhard function ${U_N(q_0, {\bf q})}$ by ${U_N(q_0-V_c(r)-Q, {\bf q})}$, where $V_{c}(r)$ is the Coulomb potential of the final nucleus. Furthermore, the renormalization of weak transition strength in the nuclear medium in a random phase approximation(RPA) is taken into account by considering the propagation of particle hole(ph) as well as delta-hole($\Delta h$) excitations~\cite{Oset2}. These considerations lead to modified hadronic tensor components $(J^{\mu\nu}_{RPA})$ involving the bilinear terms, for which expressions are given in Refs.~\cite{Nieves},\cite{Athar2}.
\begin{figure}
\includegraphics{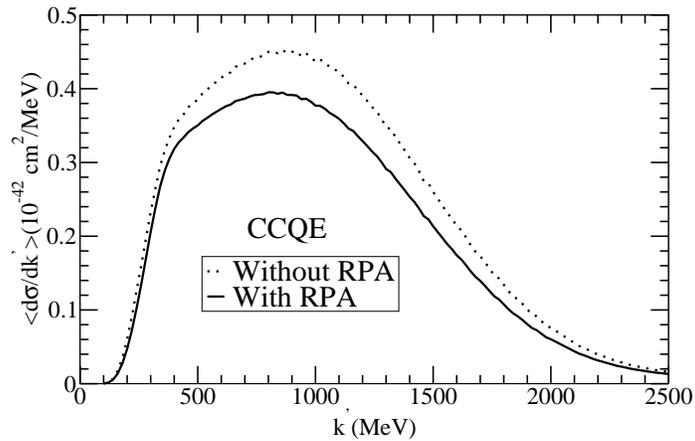}
\caption{Muon momentum distribution for $\nu_{\mu}$ induced charged current quasielastic process in $^{12}C$ averaged over the K2K spectrum.}
\end{figure}

\begin{figure}
\includegraphics{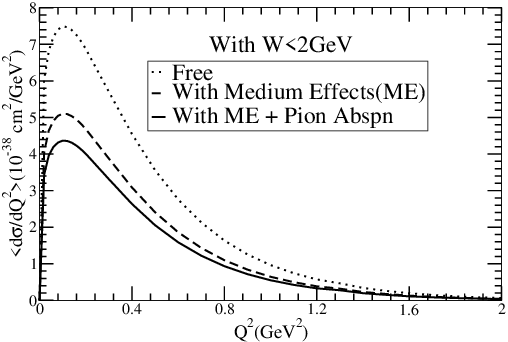}
\caption{$Q^2$ distribution for $\nu_{\mu}$ induced incoherent charged current one pion production process in $^{12}C$ averaged over the K2K spectrum.}
\end{figure} 
\section{INELASTIC RESONANCE PRODUCTION OF PIONS}
\subsection{INCOHERENT PION PRODUCTION}
The basic reaction for the inelastic one pion production in nuclei, for a neutrino interacting with a nucleon N inside a nuclear target is given by
\begin{equation}\label{inelastic_reaction}
\nu_{\mu}(k) + N(p) \rightarrow \mu^{-}(k^{\prime}) + N^{\prime}(p^{\prime}) + \pi^{+}(k_{\pi}). 
\end{equation}
In the case of incoherent pion production in $\Delta$ dominance model, the weak hadronic currents interacting with the nucleons in the nuclear medium excite a $\Delta$ resonance which decays into pions and nucleons. The pions interact with the nucleus inside the nuclear medium before coming out. The final state interaction of pions leading to elastic, charge exchange scattering and the absorption of pions lead to reduction of pion yield. The nuclear medium effects on $\Delta$ properties lead to modification in its mass and width which have been discussed earlier by Oset et al.~\cite{Oset} and applied to explain the pion and electron induced pion production processes from nuclei. 

In the local density approximation the expression for the total cross section for the charged current one pion production (kinematics is shown in Fig.1) is written as
\begin{eqnarray}\label{sigma_inelastic}
\sigma &=& \frac{1}{(4\pi)^5}\int_{r_{min}}^{r_{max}}(\rho_{p}(r)+\frac{1}{9}\rho_{n}(r)) d\vec r\int_{Q^{2}_{min}}^{Q^{2}_{max}}dQ^{2} \int^{{k^\prime}_{max}}_{{k^\prime}_{min}} dk^{\prime}\int_{-1}^{+1}dcos\theta_{\pi }\int_{0}^{2\pi}d\phi_{\pi}~~\frac{\pi|\vec  k^{\prime}||\vec k_{\pi}|}{M E_{\nu}^2 E_{l}}\nonumber\\
&&\times \frac{1}{E_{p}^{\prime}+E_{\pi}\left(1-\frac{|\vec q|}{|\vec k_{\pi}|}cos\theta_{\pi }\right)}\bar\sum \sum|\mathcal M_{fi}|^2
\end{eqnarray}
where the proton density $\rho_{p}(r)=\frac{Z}{A}\rho(r)$ and the neutron density $\rho_{n}(r)=\frac{A-Z}{A}\rho(r)$ with $\rho(r)$ as the nuclear density taken as 3-parameter Fermi Density~\cite{Vries}.
\begin{figure}
\includegraphics{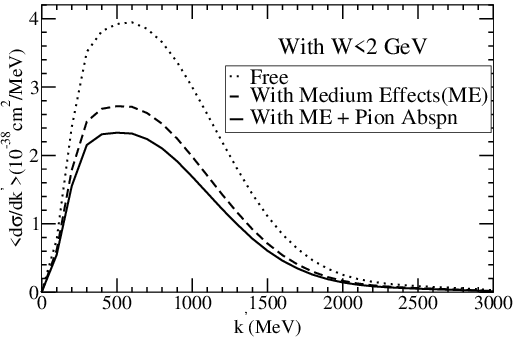}
\caption{Muon momentum distribution for $\nu_{\mu}$ induced incoherent charged current one pion production process in $^{12}C$ averaged over the K2K spectrum.}
\end{figure}

\begin{figure}
\includegraphics{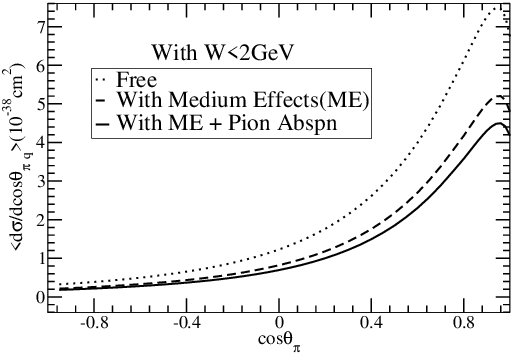}
\caption{Pion angular distribution for $\nu_{\mu}$ induced incoherent charged current one pion production process in $^{12}C$ averaged over the K2K spectrum.}
\end{figure}
\begin{figure}
\includegraphics{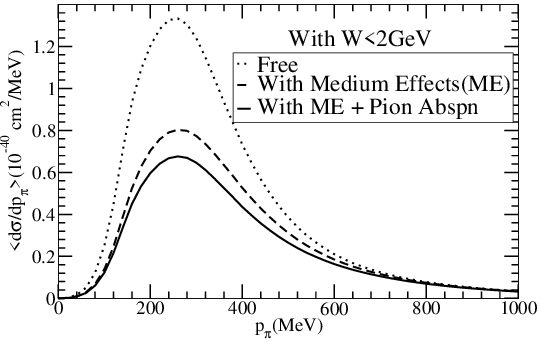}
\caption{Pion Momentum distribution for $\nu_{\mu}$ induced incoherent charged current one pion production process in $^{12}C$ averaged over the K2K spectrum.}
\end{figure}

The transition matrix element $\mathcal M_{fi}$ is given by
\begin{equation}\label{matrix_element}
\mathcal M_{fi}=\sqrt{3}\frac{G_F}{\sqrt{2}}\frac{f_{\pi N \Delta}}{m_{\pi}} \bar u({\bf p}^{\prime}) k^{\sigma}_{\pi} {\mathcal P}_{\sigma \lambda} \mathcal O^{\lambda \alpha} L_{\alpha} u({\bf p})
\end{equation}
where $L^{\alpha}$ is the leptonic current defined by Eq.(\ref{lep_curr}), and $\mathcal O^{\beta \alpha}$ is the $N-\Delta$ transition operator taken from Lalakulich et al.~\cite{Lalakulich1}. $\theta_{W}$ is the weak mixing angle. ${\mathcal P}^{\sigma \lambda}$ is the $\Delta$ propagator in momentum space and is given as~: 
\begin{equation}\label{width}
{\mathcal P}^{\sigma \lambda}=\frac{{\it P}^{\sigma \lambda}}{P^2-M_\Delta^2+iM_\Delta\Gamma}
\end{equation}
where ${\it P}^{\sigma \lambda}$ is the spin-3/2 projection operator and the delta decay width $\Gamma$ is taken to be an energy dependent P-wave decay width~\cite{Oset}:
\begin{equation}\label{Width}
\Gamma(W)=\frac{1}{6 \pi}\left(\frac{f_{\pi N \Delta}}{m_{\pi}}\right)^2 \frac{M}{W}|{\bf q}_{cm}|^3
\end{equation}
$|q_{cm}|$ is the pion momentum in the rest frame of the resonance and W is the center of mass energy.
Inside the nuclear medium, the mass and width of delta are modified which in the present calculation are taken by considering the following effects.\\
In nuclear medium $\Delta$s decay mainly through the $\Delta \rightarrow N\pi$ channel. The final nucleons have to be above the Fermi momentum $p_F$ of the nucleon in the nucleus which leads to a modification in the decay width of delta. The modified delta decay width $\tilde\Gamma$ has been taken from the works of Oset et al.~\cite{Oset} which is given by
\begin{equation}
\tilde\Gamma=\Gamma \times F(p_{F},E_{\Delta},k_{\Delta})
\end{equation}
 where 
\begin{equation}
F(p_{F},E_{\Delta},k_{\Delta})= \frac{k_{\Delta}|{{\bf q}_{cm}}|+E_{\Delta}{E^\prime_p}_{cm}-E_{F}{W}}{2k_{\Delta}|{\bf q^\prime}_{cm}|} 
\end{equation}
 $E_F=\sqrt{M^2+p_F^2}$, $k_{\Delta}$ is the $\Delta$ momentum and  $E_\Delta=\sqrt{W+k_\Delta^2}$. 

Furthermore, in the nuclear medium there are additional decay channels open
due to two and three body absorption processes like $\Delta N
\rightarrow N N$ and $\Delta N N\rightarrow N N N$ through which
$\Delta$ disappears in nuclear medium without producing a pion,
while a two body $\Delta$ absorption process like $\Delta N
\rightarrow \pi N N$ gives rise to some more pions. These nuclear
medium effects on the $\Delta$ propagation are included by modifying
the mass and the decay width of $\Delta$ in nuclear medium as
\begin{equation}
\Gamma\rightarrow\tilde\Gamma - 2 Im\Sigma_\Delta~~\text{and}~~
M_\Delta\rightarrow\tilde{M}_\Delta= M_\Delta + Re\Sigma_\Delta.
\end{equation}
The expressions for the real and imaginary part of the $\Delta$ self energy are taken from Oset et al.~\cite{Oset}:
\begin{eqnarray}
Re{\Sigma}_{\Delta}&=&40 \frac{\rho}{\rho_{0}}MeV ~~and \nonumber\\
-Im{{\Sigma}_{\Delta}}&=&C_{Q}\left (\frac{\rho}{{\rho}_{0}}\right )^{\alpha}+C_{A2}\left (\frac{\rho}{{\rho}_{0}}\right )^{\beta}+C_{A3}\left (\frac{\rho}{{\rho}_{0}}\right )^{\gamma}~~~~
\end{eqnarray}
In the above equation $C_{Q}$ accounts for the $\Delta N  \rightarrow
\pi N N$ process, $C_{A2}$ for the two-body absorption process $\Delta
N \rightarrow N N$ and $C_{A3}$ for the three-body absorption process $\Delta N N\rightarrow N N N$. The coefficients $C_{Q}$, $C_{A2}$, $C_{A3}$ and $\alpha$, $\beta$ and $\gamma$ are taken from Ref.~\cite{Oset}.

The pions produced in this process are scattered and absorbed in the nuclear medium. This is treated in a Monte Carlo simulation which has been taken from Ref.~\cite{Vicente}.
\subsection{COHERENT PION PRODUCTION}
$\nu_{\mu}$ induced coherent one pion production on $^{12}C$ target is given by $\nu_{\mu} + _{6}^{12}C \rightarrow \mu^{-} + _{6}^{12}C + \pi^{+}$ for which the basic reaction is given by Eq.(\ref{inelastic_reaction}). 
The total cross section (kinematics is shown in Fig.1) is given by 
\begin{eqnarray}\label{sigma_coherent}
\sigma &=& \frac{1}{(2\pi)^5}\int_{Q^{2}_{min}}^{Q^{2}_{max}}dQ^{2} \int^{{k^\prime}_{max}}_{{k^\prime}_{min}} dk^{\prime}\int_{-1}^{+1}dcos\theta_{\pi }\int_{0}^{2\pi}d\phi_{\pi}\frac{\pi|\vec  k^{\prime}||\vec k_{\pi}|}{8 E_{\nu}^2 E_{l}}~~\frac{M}{E_{p}^{\prime}+E_{\pi}\left(1-\frac{|\vec q|}{|\vec k_{\pi}|}cos(\theta_{\pi })\right)}\bar \sum|\mathcal M_{fi}|^2
\end{eqnarray}
where the matrix element for the reaction is given by 
\begin{equation}
\mathcal M_{fi} =\frac{G_{F}}{\sqrt{2}} cos\theta_{c} L^{\mu} J_{\mu} {\cal F}(\vec q - \vec k_{\pi})
\end{equation}

$L_{\mu}$ is the leptonic current given by Eq.(3) and $J_{\mu}$ is the hadronic current given by 
\begin{equation}
J_{\mu}= \sqrt{3} \frac{f_{\pi N \Delta}}{m_{\pi}}F(P^2) \sum _{r,s} {\bar u_{s}}(p^\prime) k_{\pi \sigma} \mathcal P^{\sigma \lambda} \mathcal O_{\lambda \mu} u_{r}({\bf p})
\end{equation}
where F($P^2$) (P=p+q) is $\Delta N \pi$ form factor given by\cite{Penner}:
\begin{equation}
F(P)=\frac{\Lambda^{4}}{\Lambda^{4}+(P^{2}-M_{\Delta}^{2})^{2}}
\end{equation}
with $\Lambda=1 GeV$.
 
${\cal F}(\vec q - \vec k_{\pi})$ is the nuclear form factor, given by 
\begin{equation}\label{ff}
{\cal F}(\vec q-\vec k_\pi)=\int d^{3}{\vec r} \left[{\rho_p ({\vec r})}+\frac{1}{3}{\rho_n ({\vec r})}\right]e^{-i({\vec q}-{\vec k}_\pi).{\vec r}}\end{equation}

When pion absorption effects are taken into account using the Eikonal approximation then the nuclear form factor ${\cal F}({\vec q}-{\vec k_\pi})$ is modified to $\tilde{\cal F}({\vec q}-{\vec k_\pi})$, which is taken to be~\cite{Athar4}:
\begin{eqnarray}\label{mod_ff}
\tilde{\cal F}({\vec q}-{\vec k_\pi})&=&2\pi\int_0^\infty b~db\int_{-\infty}^\infty dz~\rho({\vec b}, z)~J_0(k_\pi^tb)~e^{i(|{\vec q}|-k_\pi^l)z} e^{-if({\vec b}, z)}
\end{eqnarray} 
where 
\begin{equation}
f({\vec b}, z)=\int_z^{\infty} \frac{1}{2|{\vec{k}_\pi}|}{\Pi(\rho({\vec b}, z^\prime))}dz^\prime
\end{equation}
$k_{\pi}^{l}$ and $k_{\pi}^{t}$ are the longitudinal and transverse component of the pion momentum
and the pion self-energy $\Pi$ is given by
\begin{equation}\label{self_energy}
\Pi(\rho({\vec b}, z^\prime))=\frac{4}{9}\left(\frac{f_{\pi N\Delta}}{m_\pi}\right)^2\frac{M^2}{W^2}|{\vec p_\pi}|^2~\rho({\vec b}, z^\prime)~\frac{1}{W-{\tilde M}_\Delta+\frac{i{\tilde \Gamma}}{2}}
\end{equation}
with W as the center of mass energy in the $\Delta$ rest frame. 
 
\section{RESULTS AND DISCUSSIONS}
\subsection{TOTAL CROSS SECTIONS}
For the quasielastic reaction, the numerical results are obtained from Eq.(\ref {sigma_quasi}) using vector form factors $F_1(Q^2)$ and $F_2(Q^2)$ given by Bradford et al.~\cite{bradford} with  vector dipole mass ${M}_{V}$=0.84GeV and a dipole form for $F_{A}(Q^2)$ with axial dipole mass $M_A$=1.1GeV, and are shown in Fig.2. Our results for the total scattering cross section $\sigma$ without the RPA effects (dotted line) in the case of charged current quasielastic scattering is in fair agreement with the calculation done in Fermi Gas Model~\cite{Moniz},\cite{llewelyn},\cite{Gaisser} in the energy region of 0.4-3 GeV. When RPA correlations are included in our model, we see that they give rise to a reduction in the total cross section $\sigma(E)$ (solid line) which is 12-15$\%$ in the energy region of 0.4-3 GeV.  

In the case of charged current induced incoherent and coherent pion productions the results for the total cross sections are obtained from Eq.(\ref{sigma_inelastic}) and Eq.(\ref{sigma_coherent}) respectively using the N-$\Delta$ transition form factors given by Lalakulich et al.\cite{Lalakulich1} with $M_{A}$=1.1 GeV. The numerical results are shown in Fig.3 and Fig.4 for the incoherent and coherent pion productions. We see that the effect of nuclear medium in the pion production process as well as the final state interaction of pions both reduce the cross section. In the case of incoherent production of pions, the reduction due to nuclear medium effects in the production process is larger than due to final state interaction while in the case of coherent pion production, the reduction due to final state interaction is quite large as compared to the reduction due to the nuclear medium effects. However, the contribution of coherent process to the total pion production is small(5-6$\%$) in the energy region of $0.4GeV<E<3GeV$, therefore, the reduction in the total cross section is mainly given by the reduction in the cross section of the incoherent process.

In Fig.5, we show the results for the ratio of total one pion production and the quasielastic production cross sections for $\nu_{\mu}$ induced reaction in polystyrene($C_{8}H_{8}$), i.e. $R(E)=\frac{\sigma^{CC1\pi^{+}}(E)}{\sigma^{CCQE}(E)}$ (solid line) as a function of neutrino energy. A comparison with the experimental results of K2K~\cite{Rodriguez} and ANL~\cite{Radecky}, \cite{Barish} for R(E) are also shown in the figure. In this figure, we also show the ratio R(E)(dashed line) when no nuclear medium effect is taken into account either in the quasielastic process or in the inelastic process. When the cross sections are calculated without any nuclear medium modifications except the Fermi motion and Pauli principle for the quasielastic process and without any modification of $\Delta$ properties in the nuclear medium in case of inelastic process, the results for R(E) are shown by dashed-double dotted line. This should be compared with the Monte Carlo predictions used in K2K analysis\cite{Rodriguez}. We see from Figs. 2-5 that the nuclear medium modifications play an important role in quasielastic and inelastic reactions and tend to reduce the cross section ratio R(E). 

We have also studied the effect of the uncertainties in the ratio R(E) due to the use of various  vector form factors given by Budd et al.~\cite{BBBA2}, Bosted~\cite{BBBA3} and dipole form factors\cite{Galster} in the quasielastic case and the N-$\Delta$ transition form factors given by Schreiner \& von Hippel\cite{Schreiner} and Paschos et al.\cite{pas} in the inelastic case. The effect of varying the axial dipole mass $(M_{A})$ in the parameterization of axial form factor in quasielastic as well as in the inelastic processes in the region of $1.05<M_{A}<1.21$ GeV on the ratio R(E) has also been studied. These effects lead to a small change of about 4-6$\%$ in the ratio R(E).

\subsection{ANGULAR AND MOMENTUM DISTRIBUTIONS OF LEPTONS AND PIONS}
In Figs. 6 \& 7, we show the effect of nuclear medium on $Q^2$ distribution and the momentum distribution of the leptons averaged over the neutrino flux at K2K in the case of neutrino induced charged current quasielastic scattering on $^{12}C$ target. A reduction in $Q^2$ distribution in the peak region  is found to be about 30$\%$ when RPA effects are taken into account while this reduction is about 12$\%$ in the momentum distribution of leptons in the peak region. In the case of incoherent one pion production process, the $Q^2$ distribution as well as the momentum distribution of leptons are shown in Figs.8 \& 9. In these reactions charge pions are also produced and we show in Figs.10 \& 11, the angular distribution as well as momentum distribution of pions and the effect of nuclear medium on these distributions. We find that the reduction in $Q^2$ distribution in the peak region is around 30$\%$ when nuclear medium effects are taken into account which further reduces by about 14$\%$ when pion absorption effects are incorporated. For the pion momentum distribution the reduction due to nuclear medium effects is around 40$\%$ which further reduces by about 15$\%$ when pion absorption effects are taken into account. We have not shown these results for the coherent process as the contribution to the total pion production events have been found to be quite small. However, these results are interesting in their own right and will be reported elsewhere.

To conclude, we have studied in this paper the effect of nuclear medium on the inclusive quasielastic and one pion charged current production in nuclei and compared our results with the recent experimental results from the K2K experiment~\cite{Rodriguez}. The nuclear medium effects are important in this energy region and reduce the ratio by 30$\%$ around the energy region of 1GeV which becomes 20-25$\%$ in the neutrino energy region of 2-3GeV. The results for nuclear medium effects on the $Q^2$ and the momentum distribution of lepton in the quasielastic as well as inelastic reaction and the results for the nuclear medium effects on the angular and momentum distribution of pions have also been presented.
\section{ACKNOWLEDGMENTS}
One of the authors(M. S. A.) is thankful to T. Kajita and Y. Hayato (I. C. R. R., University of Tokyo) for many useful discussions and the warm hospitality provided during his stay at ICRR where part of this work was done. S. C. is thankful to the Jawaharlal Nehru Memorial Fund for the Doctoral Fellowship.

\end{document}